\documentclass[12pt]{iopart}

\usepackage{cite}
\usepackage{graphicx}
\usepackage{amssymb}
\usepackage{bm}
\usepackage[utf8]{inputenc} % Tell LaTeX that file is encoded in UTF-8
\usepackage[colorlinks=true]{hyperref}

\begin{document}

\title{Controlling carbon nanotube photoluminescence using silicon microring
resonators}
\date{\today}

\author{Adrien Noury, Xavier Le~Roux, Laurent Vivien and Nicolas Izard}
\address{Institut d'Electronique Fondamentale, CNRS-UMR 8622, Univ. Paris-Sud, 91405 Orsay, France}
\ead{nicolas.izard@u-psud.fr}

\begin{abstract}
	We report on coupling between semiconducting single-wall carbon nanotubes
	(s-SWNT) photoluminescence and silicon microring resonators. Polyfluorene
	extracted s-SWNT deposited on such resonators exhibit sharp emission peaks,
	due to interaction with the cavity modes of the microring resonators. Ring
	resonators with radius of 5~$\mu$m and 10~$\mu$m were used, reaching quality
	factors up to 4000 in emission. These are among the highest values reported
	for carbon nanotubes coupled with an integrated cavity on silicon platform,
	which open up the possibility to build s-SWNT based efficient light source
	on silicon.
\end{abstract}

\maketitle

\section{Introduction}

Among today's carbon based nanomaterials, semiconducting carbon nanotubes
(s-SWNT) stand out with their strong luminescence properties, and various
potential applications have been proposed, such as optoelectronics and
biosensing. Interestingly, s-SWNT direct bandgap could be tuned by ajusting
their chiral index, and photoluminescence (PL) of well isolated individual
s-SWNT was extensively studied, either by suspension over trenches and
pillars\cite{prl-Lefebvre, science-Chen} or encapsulation in micelle
surfactant\cite{science-Oconnell, prl-Hogele}.

PL was also observed in an ensemble of s-SWNT\cite{nano-Berger}, and more
recently in composite-like materials where nanotubes are dispersed into a
gelatin\cite{jap-Berger} or a polymer matrix\cite{prb-Zamora, ol-Gaufres},
opening the route towards device fabrication. Furthermore, it is possible to
stretch and align nanotubes dispersed in the matrix to enhance their PL
properties\cite{prb-Zamora, jpcc-Zamora}. 

However, even for polyfluorene-wrapped s-SWNT which display strong
PL\cite{ol-Gaufres, apl-Gaufres}, the quantum efficiency remains
low\cite{prb-Miyauchi}, which may put a brake on photonic applications. From
this point, it is particularly attractive to couple s-SWNT PL with optical
resonators\cite{nature-Vahala}, for PL enhancement and control of spontaneous
emission through Purcell effect\cite{pr-Purcell}.

First reports of s-SWNT PL coupling with optical resonators used planar
cavities, either using metallic mirror\cite{natnano-Xia}, dielectric
mirror\cite{ox-Gaufres}, or both\cite{apl-Legrand}.

More recent works took advantages of the Silicon-On-Insulator (SOI) platform
to couple s-SWNT PL to suspended photonic crystal cavities\cite{apl-Watahiki,
apl-Sumikura}, or silicon microdisk resonators\cite{apl-Imamura}. An advantage
of this approach is the integration with silicon photonics technology, even if
it remains challenging to efficiently couple these structures with waveguides.
On the other hand, we recently proposed an integration scheme to couple s-SWNT
PL to silicon waveguides, using evanescent wave from narrow
waveguides\cite{acsnano-Gaufres}.

We propose to go further by coupling s-SWNT with microring resonators. Silicon
microrings present various advantages for coupling with carbon nanotubes.
Intrinsically, they are traveling-wave resonators, the optical wave being
guided along the circumference of the ring, and light is efficiently loaded
and extracted with silicon waveguides, as described by the coupled mode
theory\cite{ieee-Yariv}. Moreover, the pace between two successive resonant
peaks of the cavity, or Free Spectral Range (FSR), can be easily tuned by
adjusting the ring diameter, making it easy to match s-SWNT emission
wavelengths. At last, microring resonators on SOI substrate are known to
exhibit quite high quality factor\cite{ol-Niehusmann}, turning them into an
ideal candidate for carbon nanotube integration with silicon photonics.

Here, we report on the coupling of s-SWNT PL in silicon microring resonators.
Polyfluorene-wrapped s-SWNT are deposited on top of the photonic structure to
form a very thin homogeneous layer of a few nanometers thick. We observed
sharp peaks regularly spaced and superimposed to s-SWNT broad emission peaks.
These sharp peaks were attributed to s-SWNT PL coupled to microring
resonators. We have observed quality factor as high as 4000, and we also
demonstrate tuning of the cavity resonance by changing the ring diameter.

\section{Experimental details}

\begin{figure}
	\includegraphics[width=12cm]{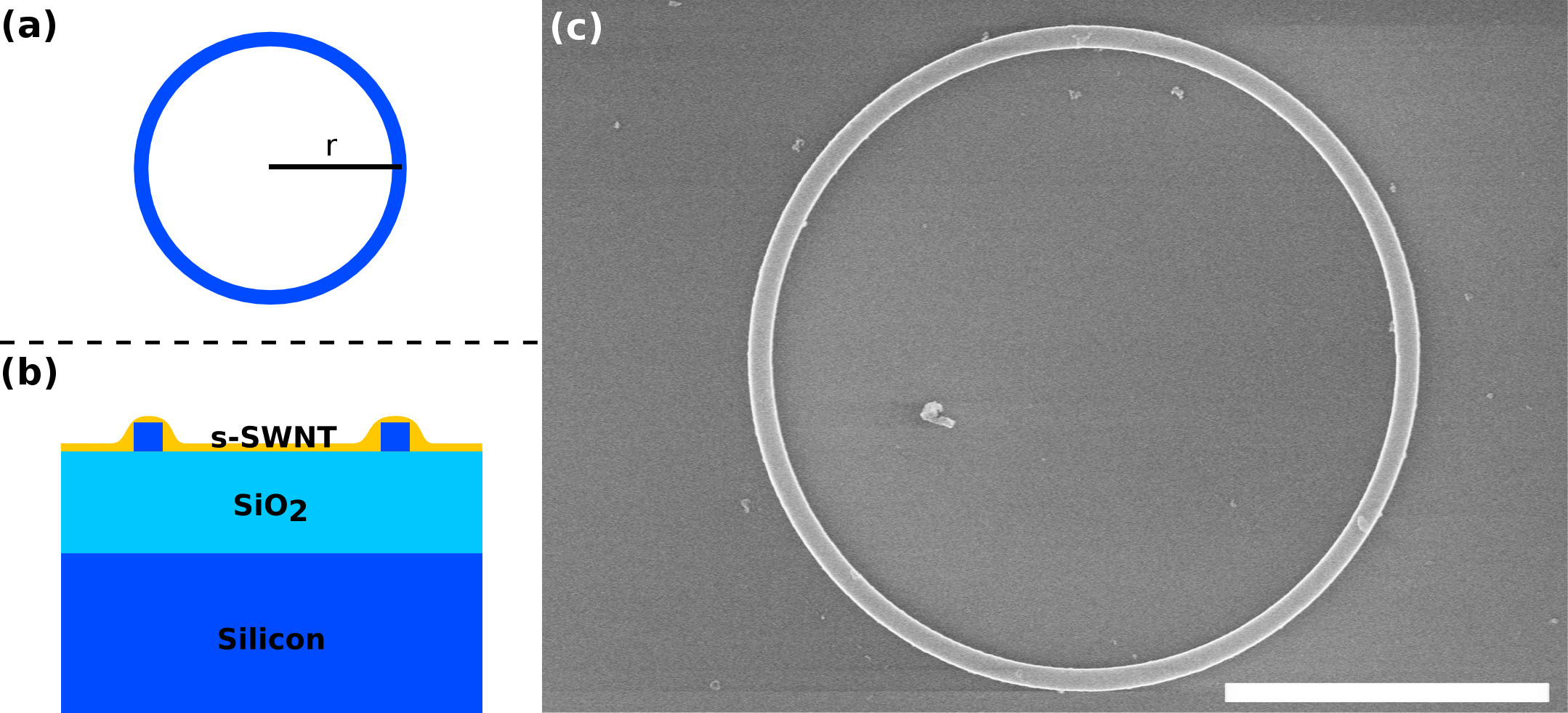}
	\caption{
		(a) Top view of a microring resonator of radius $r$. (b) Cross-section of
		the microring, with polyfluorene-wrapped s-SWNT top layer (not to scale). (c) Scanning
		electron micrograph of as-fabricated microring, with a radius $r$ of
		5~$\mu$m. The scale bar is 5~$\mu$m.
	}
	\label{fig1}
\end{figure}

The ring resonators are fabricated from SOI wafers with a 220~nm thick top Si
layer and a 2~$\mu$m thick buried oxide layer. Electron beam lithography
(NanoBeam nB4, 80~kV, 2.1~nA, step size 5~nm) is used to define the microring
and silicon waveguides. A dry etching process with an inductively coupled
plasma etcher (SF$_6$/C$_4$F$_8$) transfers the patterns in the top Si layer.
The microring resonator was made from a 350~nm wide waveguide bended to form a
circle. We used two different rings radius: 5~$\mu$m and 10~$\mu$m. The
microring was coupled to a strip waveguide of the same width, at a gap
distance $L_c$ of 85~nm between ring and waveguide. A schematic view of the
microring is shown in figures~\ref{fig1}(a) and \ref{fig1}(b), while a
scanning electron micrograph of a typical microring prior to nanotube
deposition is shown in figure~\ref{fig1}(c).

Polyfluorene-wrapped s-SWNT were prepared from carbon nanotubes produced with
the HiPCO process\cite{cpl-Nikolaev} (Unydim), using
Poly-9,9-di-n-octyl-fluorenyl-2,7-diyl (PFO) (Sigma-Aldrich) in toluene as an
extracting agent. Using ultracentrifugation up to 150000~g for 1~h, it is
known that this process leads to well isolated PFO-wrapped s-SWNT displaying
good electrical\cite{apl-Izard}, optical\cite{apl-Gaufres} and
electro-optical\cite{epjap-Izard} properties. PFO/s-SWNT are deposited by spin
casting (1000~rpm, 60~s) followed by thermal annealing of 30~min at
180$^\circ$C

\section{Results and discussions}

\begin{figure}
	\includegraphics[width=12cm]{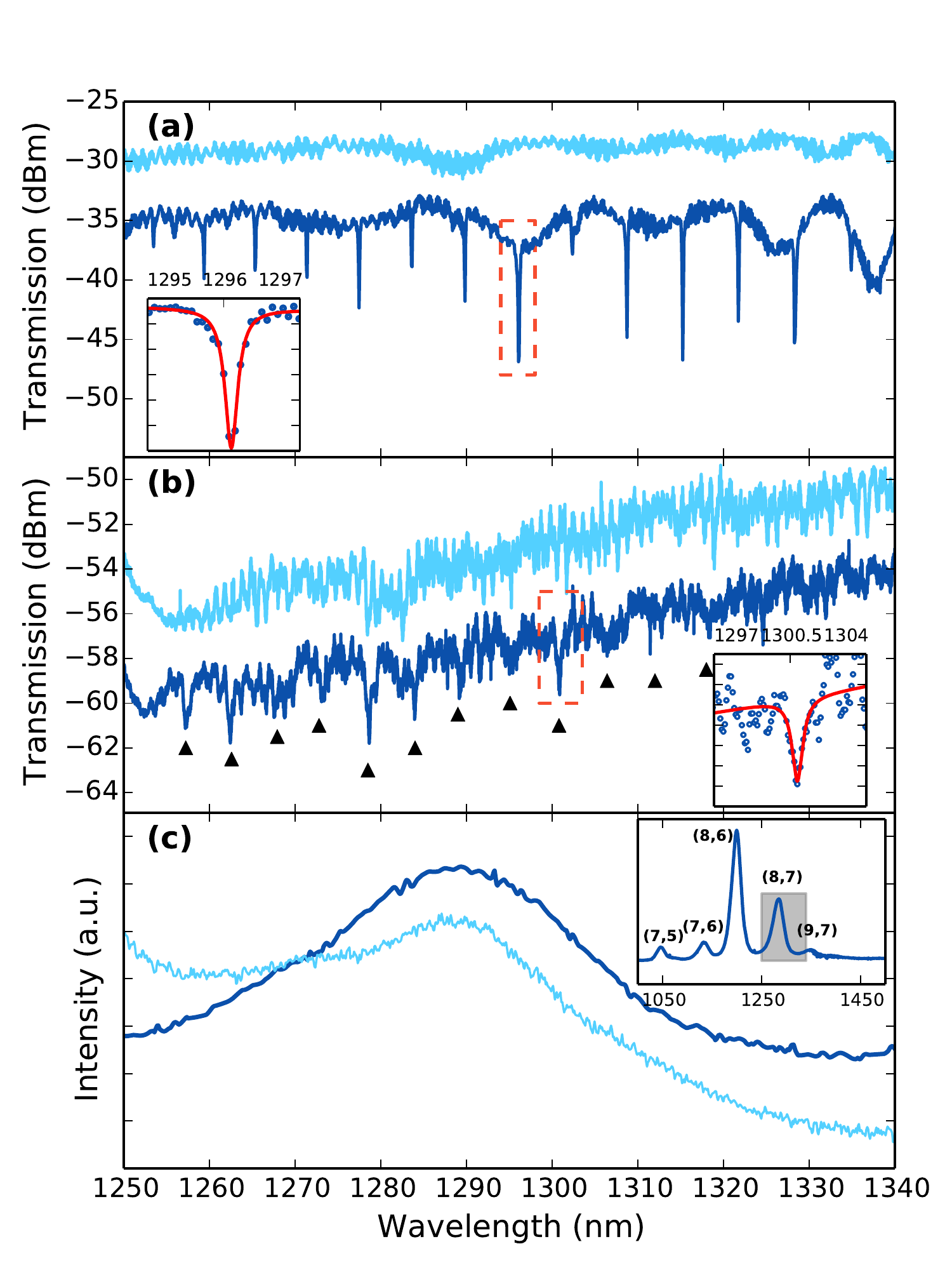}
	\caption{
		(a) Transmission spectra in TE polarization of a typical 10~$\mu$m radius microring resonator
		coupled to a strip waveguide (dark) and a reference strip waveguide
		(bright).  The reference was upshifted by 5 dBm for clarity. Depending on
		the wavelength, the distance between two resonant peaks (FSR) is comprised
		between 6 to 7~nm (see details in text). Inset depicts a close vue of one
		of the resonance, fitted with a Lorentzian lineshape (solid line). (b)
		Linear transmission of a similar microring resonator (dark) and reference
		waveguide (bright), after s-SWNT deposition. The reference was again
		upshifted by 5 dBm for clarity. Dark triangles ($\blacktriangle$) are
		guides to the eye to underline resonant peaks positions. The FSR slightly
		increases after s-SWNT deposition. Inset: detailed view of an individual
		resonance.	(c) Typical PL spectrum of s-SWNT after deposition on glass
		(dark) and SOI substrate (bright). In case of SOI substrate, the short
		wavelength region is dominated by silicon substrate PL tail. Inset shows
		the full band spectrum of s-SWNT in liquid with indexation, at an
		excitation wavelength of 735~nm.
	}
	\label{fig2}
\end{figure}

First, the fabricated microring resonators are characterized prior to s-SWNT
deposition. A strip waveguide is localized within coupling distance near to a
microring.  When light is injected into the waveguide, part of the energy will
be coupled into the ring, before being coupled back to the waveguide. For some
given wavelengths depending on the microring dimensions - radius and width of
the bended waveguide - resonances are observed, which appears as drop in
transmittance. Therefore, the transmission spectrum is characterized by sharp
absorption peaks. Figure~\ref{fig2}(a) displays a typical transmission
spectrum for a 10~$\mu$m radius microring in TE polarization. A typical
resonance is highlighted in the inset and presents a symetrical Lorentzian
lineshape. A Lorentzian fit gives a $Q$-factor of approximately 4500, for a
Full-Width Half Maximum about 290~pm. 

Silicon microring resonators are also described by their Free Spectral Range
(FSR), given by:
$$\Delta \lambda = \frac{\lambda^2}{n_g \cdot L}$$
where $\lambda$ is the wavelength, $n_g$ the group index and $L$ the length of
the optical resonator.  We found a FSR ranging from 6.18~nm at 1283.65~nm to
7.26~nm at 1383.21~nm. The corresponding $n_g$ is equal to 4.2, which is in
complete adequation with the value of $n_g$ expected for these kind of
structures, as calculated with a mode solver (not shown).

Polyfluorene wrapped s-SWNT were deposited on top of the resonators by spin
casting, and were subsequently annealed for 30~min at 180$^\circ$C. The s-SWNT
layer thickness was estimated by AFM and spectroscopic ellipsometry to be in
the order of 5~nm (not shown). Deposition of s-SWNT induces drastic changes in
microring resonators transmission as displayed in figure~\ref{fig2}(b). The
overall transmission decreases, and this is more pronounced at shorter
wavelengths.  This effect can be interpreted as, first, a change of light
confinement due to the s-SWNT layer's presence (change in the cladding
refractive index), effect which is more pronounced at lower wavelengths, and
secondly, additional absorption due to the broad (8,7) s-SWNT absorption bands
at these shorter wavelengths.

Moreover, the intensity of microring resonances significantly decreases.
Positions of these resonances are highlighted by triangle marks
($\blacktriangle$) as a visual guide. A typical resonance is displayed in the
inset. Although the intensity strongly decreases, the lineshape remains
symmetrical and can be fitted by a Lorentzian, once the general linear trend is
taken into account. The Lorentzian fit gives a $Q$-factor around 2200.
Interestingly, the positions of these resonances and the FSR slightly changes
compared to before s-SWNT deposition (FSR equal to 5.3~nm at 1283.96~nm),
which was expected for a PFO based upper cladding.

Typical PL spectra of PFO-wrapped s-SWNT are displayed in figure~\ref{fig2}(c).
The pumping wavelength was 735~nm, and s-SWNT's PL spectra were recorded on
glass (dark) and SOI (bright) substrates. The figure~\ref{fig2}(c) inset
displays PL spectrum of PFO-wrapped s-SWNT suspension in toluene for
reference. Indexation of this PL spectrum was done based on previous
works\cite{ol-Gaufres, epjap-Izard}. Due to the silicon structure bandpass,
the ($8$,$7$) s-SWNT is the main species interacting with microrings. It can
be noticed that the short wavelength region on SOI substrate is dominated by
the silicon PL tail.  However, the ($8$,$7$) emission peak is clearly
observed.

\begin{figure}
	\includegraphics[width=12cm]{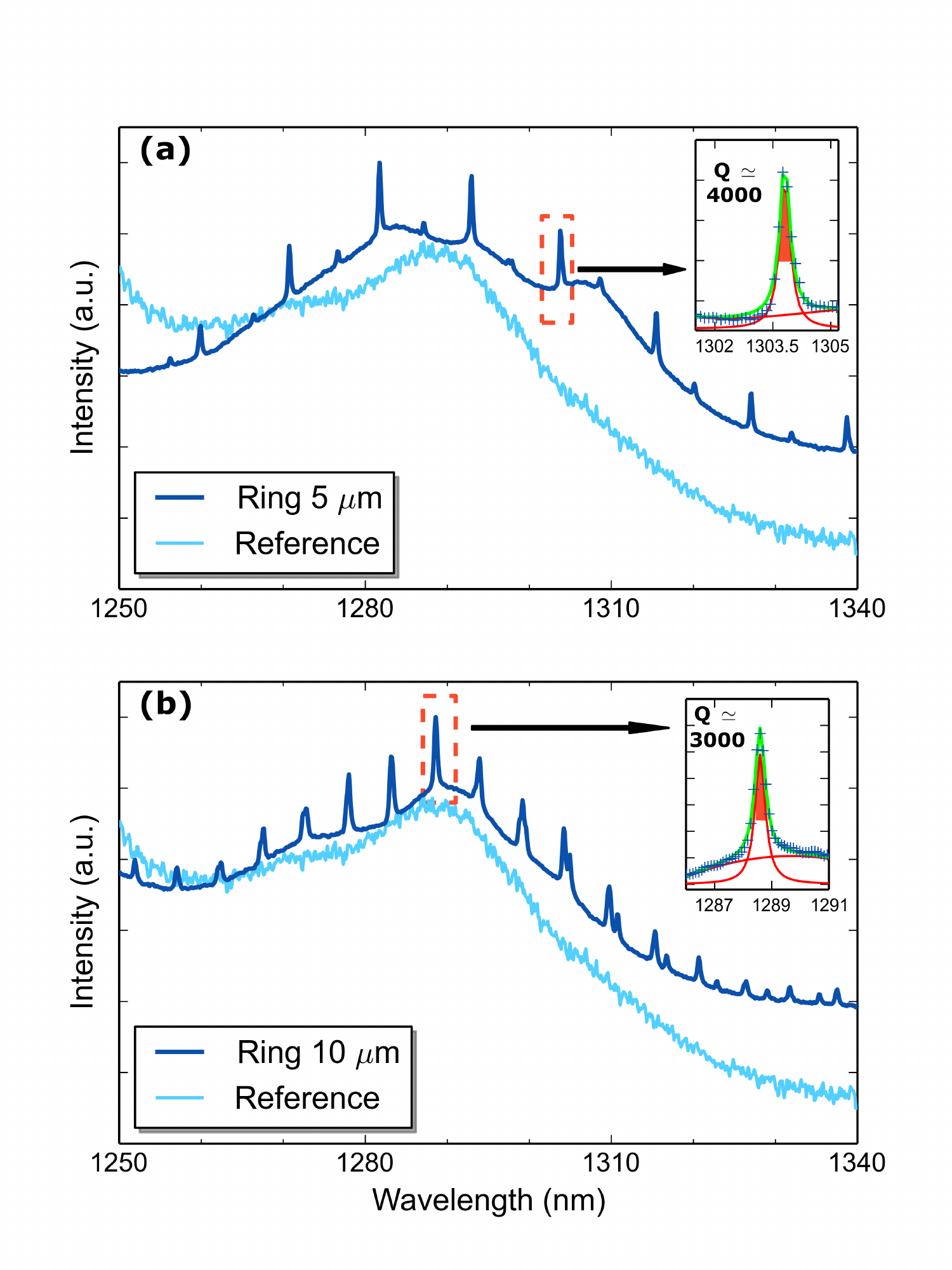}
	\caption{
		(a) Typical $\mu$-PL spectra of a 5~$\mu$m microring resonator (dark) and
		reference outside the microring (bright). (b) Same with a 10~$\mu$m
		microring. Insets show typical Lorentzian fits giving the $Q$-factor. In
		this case, FWHM is 322~pm and the center wavelength is 1303.8~nm for the
		5~$\mu$m microring, while the FWHM is 436~pm and the center wavelength is
		1288.6~nm for the 10~$\mu$m microring. Excitation wavelength is 735~nm.
	}
	\label{fig3}
\end{figure}

Using our home-built $\mu$-PL setup, s-SWNT emission on top of microring
resonators is measured. The output of a titanium-sapphire laser at 735~nm is
focused onto the sample with an infrared microscope objective of 0.6 numerical
aperture. PL is collected by the same objective, and the direct laser line is
rejected by a dichroic beam splitter and a longpass filter. PL is recorded
with a 320-mm spectrometer and 950 lines/mm grating, and detection is
performed by a nitrogen-cooled 512-pixel linear InGaAs array. In this
configuration, spot size diameter on top of silicon microrings was 15~$\mu$m,
and the excitating power was maintened below 5~mW.

Typical results for 5 and 10~$\mu$m radius ring resonators are displayed in
figures~\ref{fig3}(a) and (b), respectively. The most striking feature is the
appearance of sharp peaks regularly spaced, superimposed to the ($8$,$7$)
s-SWNT broad emission peak. Their quality factors $Q$ range mostly from 3000
to 4000. These sharp peaks are attributed to nanotube PL coupled to the
microring whispering gallery mode. Part of the light emitted by the nanotube
is directly collected by the microscope objective, resulting in the broad
emission peak. However, a significant fraction of the energy couples to the
microcavity modes: at resonant wavelengths, light emitted by the s-SWNT loads
into the microring cavity. Intrinsic resonator losses induce far-field
radiation, resulting in enhanced emission at these wavelengths compared to off
resonance s-SWNT emission.

\begin{figure}
	\includegraphics[width=12cm]{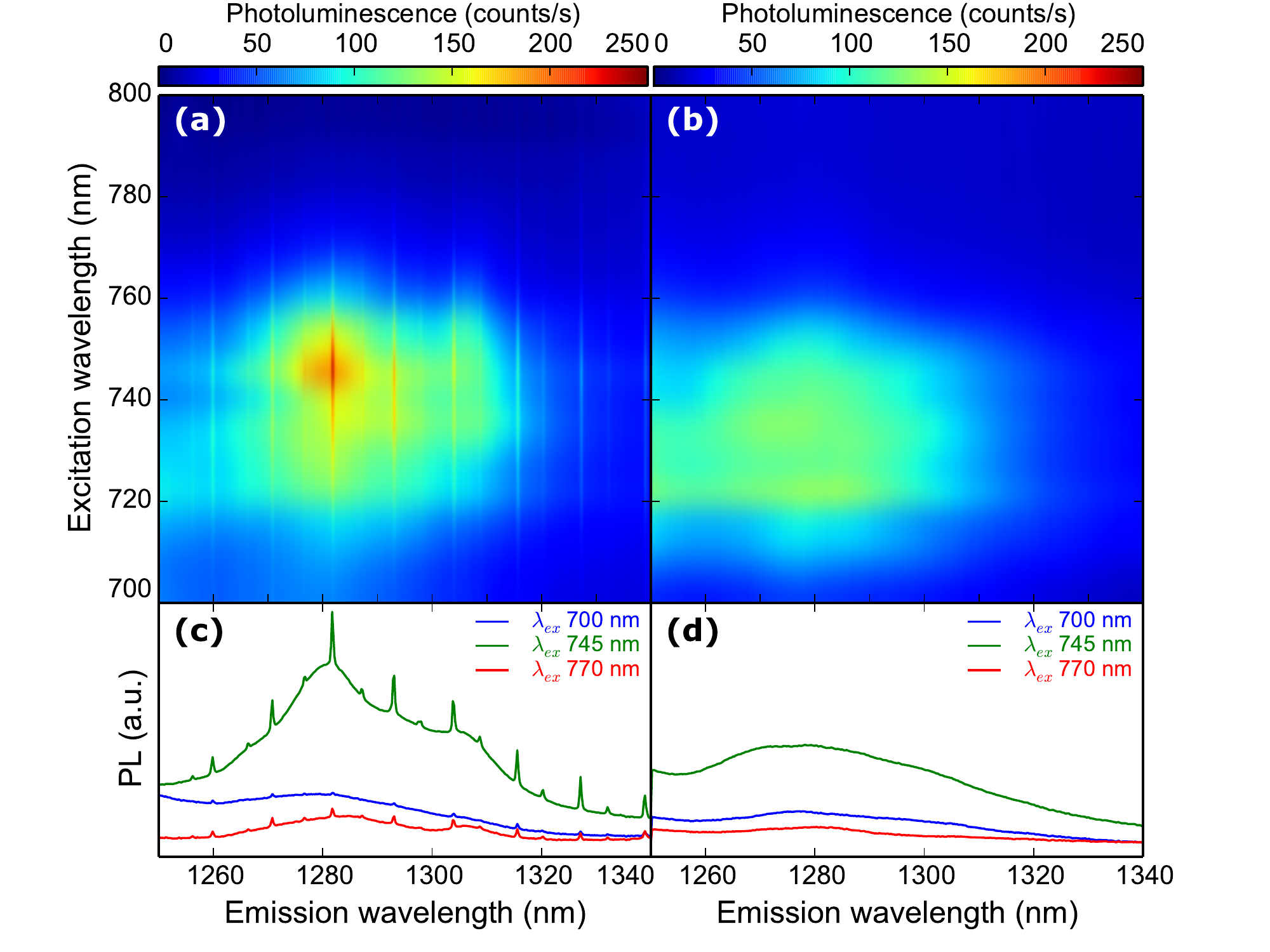}
	\caption{
		Top[(a), (b)] PLE map, and bottom[(c), (d)] corresponding PL spectra at
		different wavelengths for, respectively, a 5~$\mu$m microring
		resonator [left (a), (c)] and the reference substrate [right (b), (d)].
		The peak centered around 1290~nm is attributed to (8,7) s-SWNT, while the
		signal on the edge at 1250~nm is attributed to the (8,6) and Si PL tail.
	}
	\label{fig4}
\end{figure}

As the FSR directly depends on the microring radius, it is therefore expected
that if the microring radius doubles (e.g. from 5~$\mu$m radius to 10~$\mu$m
radius), the FSR will be halved. In figure~\ref{fig3}(a) the FSR around 1290~nm
is between 10 to 11~nm, while in figure~\ref{fig3}(b), the FSR is equal to
5.3~nm, which unambiguously proves that these sharp peaks can be attributed to
microring resonances (with the FSR being of the same order as determined by
linear transmission).

Moreover, the intensity of the microring peaks is modulated by the broad
s-SWNT PL, the intensity of the peak being higher when the SWNT peak intensity
is high. (cf. Supplementary Materials). This is another demonstration that
these resonance peaks originate from s-SWNT PL coupled into the microring
resonators.

For the 5~$\mu$m and 10~$\mu$m microring resonator, an additional serie of
peaks is observed, with a lower intensity and a different FSR than the main
serie. Although it is difficult to unambiguously assign these peaks, they
might origin from TE/TM polarization, the low intensity set being attributed
to TM modes.

To further investigate the coupling between s-SWNT and microring resonators,
photoluminescence excitation (PLE) spectroscopy was performed (cf.
figure~\ref{fig4}). A striking feature of PLE map performed on top of
microring resonators is the presence of sharp vertical peaks, disappearing
when the (8,7) s-SWNT PL intensity drop. This effect is clearly unobserved
when PLE map is performed outside the microring (figures~\ref{fig4}(b) and
(d)). These results again demonstrate that s-SWNT PL is well coupled into
silicon microring, resulting in sharp emission peaks at the microring
resonance.

\section{Conclusion}

In conclusion, we have demonstrated the coupling of s-SWNT PL in silicon
microring resonators, resulting in sharp and regularly spaced emission peaks.
The Free Spectral Range of these emission peaks could be easily tuned by
adjusting the microring diameter. The quality factor $Q$ of these emission
peaks range from 3000 to 4000, which is among the highest values reported for
integrated silicon cavities coupled to carbon nanotubes. Our results open up a
range of possibilities for future photonic circuits, due to the straightforward
coupling between silicon microring and strip waveguide\cite{ol-Niehusmann} and
their integration into more complex devices. For instance, recent results
suggest that carbon nanotubes network could be electrically-driven to produce
light in telecom wavelengths range, light originating from black body
emission\cite{apl-Fujiwara}. Coupled with silicon microring resonators into
more complex devices, this could be the first step towards the implementation
of on-chip light multiplexing. Additionally, enhanced localization of light in
photonic crystal can increase significantly the interaction between s-SWNT PL
and the propagating optical modes, leading to strong photoluminescence signal
in such devices\cite{APL-Charles}.

\section*{Acknowledgments}

A. Noury acknowledges the Ministry of Higher Education and Research (France)
for scholarship. The authors acknowledge C. Caër and A. Degiron for assistance
and fruitful discussions. This work was supported by ANR JCJC project
"\c{C}a~(Re-)~Lase~!" and partly by the FET project "Cartoon". Fabrication was
performed in the IEF clean room facilities (CTU/MINERVE), part of the RENATECH
network.

\section*{References}

\bibliography{microring}{}
\bibliographystyle{unsrt}

\end{document}